\documentclass[aps,prl,twocolumn,superscriptaddress,amsmath,amssymb]{revtex4-1}
\usepackage[latin1]{inputenc}
\usepackage{bm}
\usepackage{color, amssymb, graphicx, amsfonts,epstopdf }
\usepackage{multirow,amssymb,amsbsy,amsmath,epstopdf}
\usepackage{graphicx}
\usepackage{verbatim}
\newcommand{\bra}[1]{\ensuremath{\left\langle #1\right|}}
\newcommand{\ket}[1]{\ensuremath{\left|#1\right\rangle}}
\newcommand{\braket}[2]{\ensuremath{\left\langle #1|#2\right\rangle}}

\begin{document}

\title{Experimental observation of anomalous trajectories of single photons}

\author{Zong-Quan Zhou}
\author{Xiao Liu}
\affiliation{Key Laboratory of Quantum Information, University of Science and Technology of China,
CAS, Hefei, 230026, China}
\affiliation{Synergetic Innovation Center of Quantum Information and Quantum Physics, University of Science and Technology of China, Hefei, 230026, China}
\author{Yaron Kedem}\email{yaron.kedem@fysik.su.se}
\affiliation{Nordita, Center for Quantum Materials, KTH Royal Institute of Technology and Stockholm University, Roslagstullsbacken 23, 10691 Stockholm, Sweden}
\author{Jin-Min Cui}
\author{Zong-Feng Li}
\author{Yi-Lin Hua}
\author{Chuan-Feng Li}\email{cfli@ustc.edu.cn}
\author{Guang-Can Guo}
\affiliation{Key Laboratory of Quantum Information, University of Science and Technology of China,
CAS, Hefei, 230026, China}
\affiliation{Synergetic Innovation Center of Quantum Information and Quantum Physics, University of Science and Technology of China, Hefei, 230026, China}
\date{\today}

\begin{abstract}
A century after its conception, quantum mechanics still hold surprises that contradict many ``common sense'' notions. The contradiction is especially sharp in case one consider trajectories of truly quantum objects such as single photons. From a classical point of view, trajectories are well defined for particles, but not for waves. The wave-particle duality forces a breakdown of this dichotomy and quantum mechanics resolves this in a remarkable way: Trajectories can be well defined, but they are utterly different from classical trajectories. Here, we give an operational definition to the trajectory of a single photon by introducing a novel technique to mark its path using its spectral composition. The method demonstrates that the frequency degree of freedom can be used as a bona fide quantum measurement device (meter). The analysis of a number of setups, using our operational definition, leads to anomalous trajectories which are non-continuous and in some cases do not even connect the source of the photon to where it is detected. We carried out an experimental demonstration of these anomalous trajectories using a nested interferometer. We show that the Two-state vector formalism provides a simple explanation for the results.
\end{abstract}
\maketitle

\section{I. INTRODUCTION}

Ever since the days of Einstein and Bohr, quantum mechanics has incited heated debates. While its predictions are largely undisputed, there is no consensus on what this theory can tell us about the past. This problem is usually formulated by asking in ``which way" a particle had passed inside an interferometer. In recent years, as the advancements in technology have opened new experimental possibilities, there has been growing interest in this problem. Understanding the behavior of single particles is not only important for fundamental considerations, but also useful for generating novel information processing protocols \cite{app}. Much of the work was in the context of complementarity \cite{comp1,comp2}, where obtaining information regarding the path of a particle is traded for the quality of interference. In other work, the counter intuitive nature of the theory was used to conceive new concepts, with a potential for practical applications, such as counter factual computing \cite{counter} and the quantum eraser \cite{eraser}.

A different approach to this problem is to look on the limit of vanishing ``which way'' information, and near perfect interference, in order to study the past of an undisturbed quantum system. The past of a classical particle is described by a trajectory and this concept was extended to quantum particles \cite{bohm,bohm2,BB} and to waves \cite{BL,Prosser}. In the case of classical waves, trajectories can be connected to the flow of energy or momentum and one can describe the evolution of a system using a trajectory equation. However, the inherent non-local nature of waves distinguishes the wave trajectories from ones related to a single particle. In the quantum case, the trajectory of particles can be defined via Bohmian mechanics \cite{bohm}, which include the surreal case \cite{mahler}, where experimental observations are not indicative to the particle location, and also manifest superluminal effects.

In this work, we start by defining an operational trajectory, based only on experimental observations and independent of a microscopic theory or interpretation. It would be desirable for any microscopic definition for the trajectory to agree with the operational one. Unfortunately, the results show that in some scenarios the operational trajectory are not continuous, a fact that sets tough restrictions on any microscopic description. A description that satisfies these restrictions and indeed agree with the results in the scenarios we studied, comes from the two-state vector formalism (TSVF) \cite{TS}. The system is described by a forward evolving quantum state $\ket{\psi}$, representing the preparation of the system, and by a backward evolving quantum state $\bra{\phi}$, representing the state in which the system was found finally. The trajectory is defined as any place where both $\ket{\psi}$ and $\bra{\phi}$ have finite support at some time.

\section{II. OPERATIONAL DEFINITION of PHOTONIC TRAJECTORY}

Our method is inspired by Weak measurements \cite{AAV}, a technique that has been shown to be highly useful for investigating such issues \cite{,traj,wavefunction} and for practical applications as well \cite{snr,x1,x2,led}. However, our definition of operational trajectories does not depend on any result related to weak measurement and one does not require weak values in order to understand our method. The key issue is that the impact of the measurement process on the system is controlled by a small parameter, so the limit of vanishing impact is well defined.

A recent proposal by Vaidman \cite{Vaidman13} and its implementation by Danan \emph{et al.} \cite{Danan13}, have utilized a similar approach. Their observations limit drastically the possibilities for considering the past of a particle. In particular, it rules out the option of a continuous trajectory. The results have stirred a lively discussion in the literature \cite{viewpoint,Li13,Vaidman13b,Salih13,Danan15,Bartkiewicz15,Potocek15,Vaidman16,dove,ideal,classic,tracing} including criticism on the use of classical light and the sensitivity of the experiment. Unfortunately, the clever technique used for obtaining the ``which path" information, could not be applied to single photons. The use of a split-detector means that the dichotomic answer from a single photon cannot be directly related to the ``which path" question. Only when comparing the answers of photons at different times the information can be obtained.  Thus, the results could be interpreted as coming from classical waves, for which the notion of trajectories is not suitable anyway. Obtaining a similar result for single photons is non-trivial. For example, if one replace the split-detector with a continuous position detector and the vibration of different frequency with a variable deflection, the result is a sum of the deflections (in the photons trajectory) and the uncertainty excludes an operational definition for a single photon. One has to record, in the state of the photon, some information regarding the path that can be obtained conclusively from a single photon, but without destroying the interference. The fundamental nature of this obstacle has strong implications on interpreting the trajectories. If one cannot devise a way to obtain the results with single photons, it might be that the conclusion is not valid for that case. Moreover, if different results are obtained, the conclusion might be plain wrong. Here, we overcome this obstacle and use single photons to reconstruct the trajectories. The obtained results that are similar to that obtained with classical light in some cases and different for others.

We solve this problem by utilizing the spectral composition of the photons which is controlled by the Electro-optic Phase Modulators (EOM). Our method yields distinct features for a single photon and by that provides us an operational definition of its trajectory, which are different, in some scenario, from the case of classical waves \cite{Danan13}. We show, for three different setups, how the experimentally observed trajectories, while contradicting a ``common sense'' approach, agree with the intuition provided by the TSVF.

In order to show that the resulting spectral features can be used as an operational definition of a trajectory, let us examine the way this device affects the spectrum of a photon. The light passes a region of length $L$ in which the refractive index changes as a function of time $n(t) = n_0 + g {c \over L} \sin (\Omega t)$ where $\Omega$ is the modulation frequency, $n_0$ is the time independent refractive index, $c$ is the speed of light in vacuum and $g$ is the modulation strength having dimension of time. The refractive index does not change much in the time it takes for light to pass the device, $L/c \ll \Omega^{-1} $, but it does during the coherence time of the photon $\tau \gg \Omega^{-1}$.
A component of the wave packet, which passes the EOM at $t$, will acquire a relative phase of $ e^{-i \omega g \sin (\Omega t)}=1 + \omega g \left(e^{-i \Omega t} - e^{i \Omega t}\right)/2 + O(\omega g)^2$, where $\omega$ is the  frequency of the light. The factors $e^{\pm i \Omega t}$ induce a translation in the frequency. Thus a wave packet $\Psi(\omega)$ of a single photon going through the EOM will be transformed
\begin{equation}\label{trans}
\Psi(\omega) \rightarrow  \Psi'(\omega) = \Psi(\omega) + \omega g \Psi(\omega \pm \Omega) + O\left( \omega g\right)^2.
\end{equation}
The modified wave packet is then a superposition of the original wave packet and one with a frequency shift, given by the frequency of modulation. If a photon, which initially has some central frequency $\omega_0$ and spectral width $\Delta \omega = 2 \pi / \tau \ll \Omega$, is later found with frequency close to $\omega_0 \pm \Omega$, it must have passed through the EOM. Thus, the spectral information allows us to build a trajectory of a single photon, simply by looking on the experimental result. This is our operational definition for the trajectory.

Let us now show that this procedure is also a novel method for performing Weak Measurements using the frequency of a photon as a quantum meter. Some experiments \cite{led,dopler,atom} have used weak measurements in the context of time/frequency measurements but with imaginary Weak Values so the meter could be seen as a classical variation \cite{timeWV}. When $\omega g \ll 1$ the modified wave packet can still overlap with the original one, or with a wave packet modified by a different $\Omega$, so interference is possible. Now consider splitting a wave packet, having initially some spectrum $\Psi_I(\omega)$, into two channels $\ket{\psi} = \alpha \ket{0} + \beta \ket{1} $, where $\alpha$ ($\beta$) is the amplitude in channel $\ket{0}$ ( $\ket{1} $). The light passes an EOM, only in channel $\ket{1} $, and then the two channels are joined as $\ket{\phi} = \gamma \ket{0} + \delta \ket{1} $, i.e. channel $\ket{0}$ ( $\ket{1} $) has a transfer amplitude $\gamma$ ($\delta$) to a chosen port. The spectrum of the photon coming out of that port is given by
\begin{equation}\label{meter}
\Psi_f(\omega) = \Psi_I(\omega) + (P_1)_w \omega g \Psi_I(\omega \pm \Omega) + O\left( \omega g\right)^2
\end{equation}
where $(P_1)_w = {\bra{\phi} P_1 \ket{\psi} \over\braket{\phi}{\psi}}$ is the Weak Value of the projection operator $P_1 = \ket{1} \bra{1}$. Since the weak value determines the final state of the meter, we can extract it from the spectral information of the photon. This fact allows us to analyze our system using Weak Values and it also implies that procedure is a genuine weak measurement in the sense that it demonstrates the typical phenomena associated with this technique.

\begin{figure}
\begin{center}
\includegraphics [width= 1 \columnwidth]{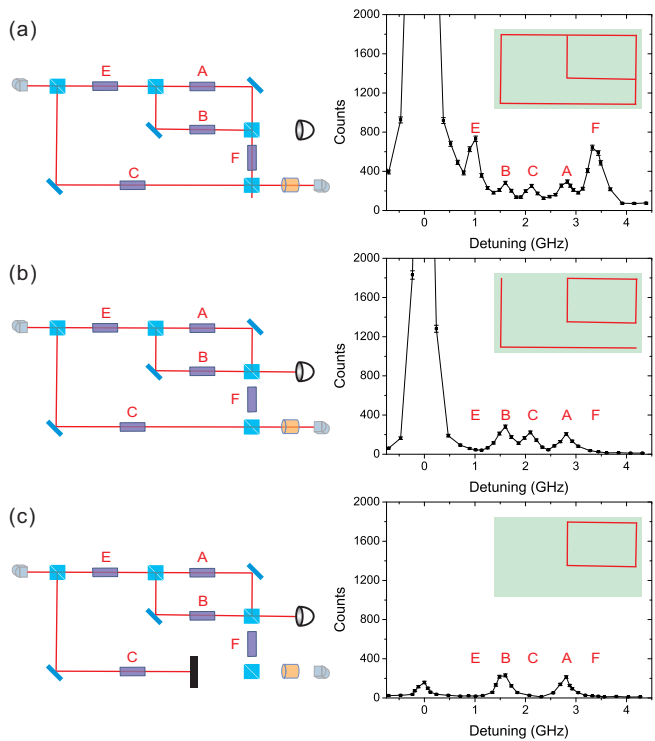}
\end{center}
\caption{(Color online) Single photons enter a nested Mach-Zehnder interferometer with EOMs, having different modulation frequencies, placed in each section. That means that in each optical path connecting two beam splitters the photon is affected by a different frequency and this information is imprinted in its state. At a chosen exit port the photon reach a tunable frequency filter so the presence of a specific component in its spectrum is recorded. The spectrum is shown in the plots, where the labels (A,B,C,E,F) show the modulation frequency according to the labels in the scheme of the interferometer. The modulation frequency for A,B,C,E and F is 2.8 GHz,1.6 GHz, 2.1 GHz, 1.0 GHz and 3.4 GHz, respectively. The experiment is performed in different configurations:(a), The inner interferometer is tuned to constructive interference (toward F). The measured frequency spectrum shows peaks at frequencies of all the EOMs inside the interferometer. (b), The inner interferometer is tuned to destructive interference. The frequency spectrum still shows peaks at the frequencies of the two EOMs inside the inner interferometer, A and B, and the one in the outer arm C, but none at the frequencies of E nor F. (c), The outer arm is blocked. Only the two peaks corresponding to the two EOMs inside the inner interferometer remain. The central peak is vertically clipped in (a) and (b). The trajectories, as defined by the TSVF, are shown as insets in each plot.}
\label{main}
\end{figure}

A schematic of the experimental setup is shown in Fig. \ref{main} (see Fig. \ref{setup} for details). Single photons enter a nested Mach-Zehnder interferometer with an EOM on each arm and their frequency is measured at the exit port. Let us give a pictorial description, which is still faithful to the technical details of the experiment: We send photons, one by one, through a network of routes and if they come out of a specific port we question them regarding the path they took. However, we are only allowed one yes/no question for each photon. The question is whether it experienced a modulation frequency $\Omega$ along its route. That is, in each run we choose a specific $\Omega$ and get a yes/no answer, or no answer if the photon came out of another port. In the plots shown in Fig. \ref{main}, the horizontal axis represents the question we choose to ask, i.e. $\Omega$, and the vertical axis represents the number of times the answer was positive. For convenience we put labels on the modulation frequency of each EOM according to their labels in the schematic of the setup.

\section{III. DETAILED EXPERIMENTAL SETUP}

\begin{figure*}[t]
\begin{center}
\includegraphics [width= 5 in]{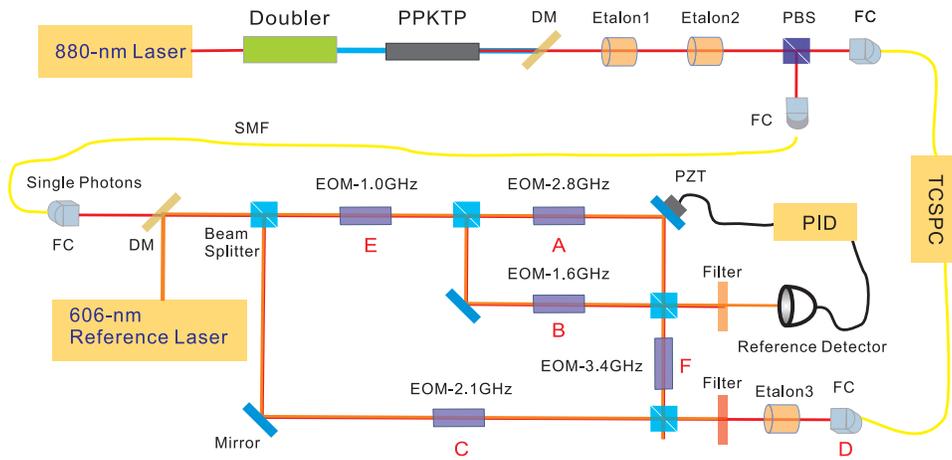}
\end{center}
\caption{(Color online) Detailed experimental setup. Light from a 880-nm laser enter a cavity-enhanced frequency doubler. The resulting blue light is sent into PPKTP crystal, acting as a pump for the down conversion process and then removed by a dichroic mirror (DM). The degenerate photon pairs, created in the crystal, are spectrally filtered by etalon1 (free spectral range of 105 GHz, and linewidth of 1.4 GHz) and the etalon2 (free spectral range of 22 GHz, and linewidth of 315 MHz). Each pair is separated by a polarization beam splitter (PBS). The horizontally-polarized photon are directed to a single-mode fiber (SMF)-coupled single-photon detector (SPD), heralding the vertically-polarized photon, which is directed to the interferometer. The inner interferometer is stabilized by measuring the interference of a 606-nm laser and feedback into a PZT-driven mirror. The 606-nm laser and 880-nm single photons are separated with interference filters. The spectrum of the photon exiting the interferometer is measured by etalon3 (free spectral range of 8 GHz, linewidth of 100 MHz). The single-photon signals are analyzed with time-correlated single-photon-counting-systems (TCSPC).
}
\label{setup}
\end{figure*}

The detailed experimental setup is shown in Fig. \ref{setup}. The 880-nm laser is a Ti:Sapphire laser (MBR-110, coherent). The 606-nm laser is a frequency doubled diode laser (Toptica, TA-SHG). The frequency of both lasers are stabilized to low-drift Fabry-P\'{e}rot Interferometers. The linewidths of both laser are well below 50 kHz. Spontaneous parametric down-conversion (SPDC) in a PPKTP crystal followed by frequency filtering using two etalons, yields degenerate pairs of photons centered at 340696.55 GHz frequency. The filtered photons have a single longitudinal mode with bandwidth of approximately 315 MHz \cite{zhouLGMM}. The measured cross-correlation between the photon pair indicates that the heralded signal photon is close to ideal single photon source \cite{lgi,entangle}. The power of the 440-nm pump light for SPDC is maintained at approximately 20 mW and the integration time is 20 minutes for measurement of each data point.

The beam splitters (BS) in the interferometer are constructed using waveplates and polarization beam splitters. The first BS and the final BS have a reflection/transmission (R/T) ratio of 1:2 and 2:1, respectively. The other two BS have R/T ratio of 1:1. Each path inside the interferometer is equipped with an free-space EOM with specific working frequency as indicated in Fig. 2. The modulation strengths of these EOM are small so that $\omega g\simeq0.025$. The higher order harmonics can be also created inside the EOM with amplitude of the order of $(\omega g)^2$, which can result into photon counts of roughly the size of the error bar in our plots. All the EOM are resonant EOM with bandwidth of approximately 1 MHz.

The phase difference of the inner interferometer can be controlled and stabilized through a PZT-driven mirror. The inner interferometer can achieve a visibility of 99.0\% for the 880-nm single photons with zero frequency detuning. The phase in the external interferometer was not crucial since it can only affect the magnitude of the peaks in the spectrum rather than their presence. It was manually tuned before each experiment and free running during the experiment.

The spectrum of single photons is analyzed with etalon3. We measured the resonant frequency of this etalon depending on the temperature. The change of resonant frequency $\Delta\nu$ shows linear dependence on the change of temperature $\Delta$T,  $\Delta\nu/\Delta$T$=-2.358\pm0.003$ (GHz/Kelvin). The temperature drift of the etalon3 is below 4 mK, which translates to a negligible frequency drift of approximately 10 MHz. The spectrum resolution in the experiment is approximately 50 MHz, which is primarily determined by the linewidth of etalon3.

\section{IV. RESULTS AND DISCUSSION}

The distribution of the results depends on the interferences between the different paths. We start by tuning the inner interferometer to a constructive interference towards F, as seen in Fig. \ref{main} (a). We obtain photons with all the relevant modulation frequencies, as seen in the plot. When we tune the inner interferometer to a destructive interference towards F, as seen in Fig. \ref{main} (b), we get a surprising result. We find photons that experienced modulation frequencies A, B and C, but practically none that experienced modulation frequencies, E and F. This result clearly contradicts the notion of a continuous trajectory and it is intensified by the fact that in each run of the experiment there was only one photon in the interferometer.

We further look on another configuration, which consist of blocking the outer arm of the interferometer, as shown in Fig. \ref{main} (c). The result is that only two frequencies are observed: A and B. This is different from what was observed in \cite{Danan13}, where the frequencies were well defined only for an ensemble of photons. Since our method rely on obtaining information from a single photon, the resulting trajectory is not only non-continuous but does not even allow the photon to travel from its source to where it was detected.
This surprising result also has a natural explanation in terms of the TSVF \cite{Salih13}. Looking on where both the forward evolving quantum state $\ket{\psi}$ and the backward evolving quantum state $\bra{\phi}$ do not vanish, we find that indeed the photon was present exactly in the regions where the EOM A and B were located. Analyzing this scenario using weak values is not straightforward since the orthogonality of the forward and backward wave functions $\braket{\phi}{\psi} = 0$ implies the weak values diverge. Thus one cannot neglect terms $O\left( \omega g\right)^2$ in Eq. (2) since they might be  $\sim 1 / \braket{\phi}{\psi} $.

The issue can be resolved by considering the inner interferometer to be imperfect, a reasonable assumption. We write the forward evolving quantum state as $\ket{\psi}= {1 \over \sqrt{3}} \left( \ket{C} + \ket{A} + e^{i \epsilon} \ket{B} \right)$ and the backward evolving quantum state as $\ket{\phi}= {1 \over \sqrt{2}} \left(  \ket{A} - \ket{B} \right)$, where $\ket{A}$,  $\ket{B}$, $\ket{C}$ are the paths going through EOM A, B, C respectively and $\epsilon \ll 1$ is the imperfection of the inner interferometer. For projections on regions outside the interferometer, $P_E = {1 \over 2} \left( \ket{A} + e^{i \epsilon} \ket{B} \right) \left( \bra{A} + e^{-i \epsilon} \bra{B} \right)$ and $P_F = {1 \over 2} \left( \ket{A} - \ket{B} \right) \left( \bra{A} - \bra{B} \right)$, the weak values are $(P_F)_w =  (P_E)_w =1$. For projections on regions inside the interferometer, $P_A = \ket{A}  \bra{A}$ and $P_B = \ket{B}  \bra{B}$, the weak values are  $\left| (P_A)_w \right| \sim  \left| (P_B)_w \right| \sim \epsilon^{-1}$. As can be seen from Eq. (\ref{meter}) the change in the spectrum is of order $g (P_X)_w$, where $X = A,B,E,F$. Taking the limits $\epsilon \rightarrow 0, g \rightarrow 0$ while keeping ${g \over \epsilon}$ finite, we conclude that only the regions inside the interferometer should have impact on the meter. The conclusion is in agreement with the simple picture of the TSVF and also with our experimental results. This is an additional support for the TSVF and it is revealed by the sensitivity of our setup which is required due to the use of single photons.

\begin{figure}
\begin{center}
\includegraphics [width= 0.8 \columnwidth]{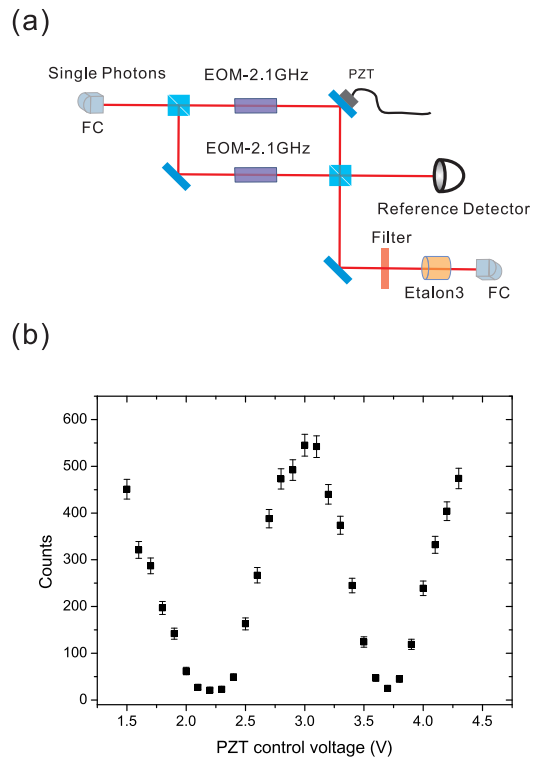}
\end{center}
\caption{(Color online) (a). Two EOMs with the same modulation frequencies of 2.1 GHz are placed in two arms of an interferometer. The two EOM are driven with a single microwave source to avoid a phase difference introduced by the microwave. Only the frequency-shifted components are measured after the etalon. (b). By tuning the control voltage of PZT, one can see near-perfect interference. The achieved visibility is approximately 97.6\% which is very close to the visibility for the components other than the EOM.}
\label{second}
\end{figure}

Since the inner workings of the EOM are not so transparent, one might question the validity of our quantum treatment to what happens inside this device. In particular, it is important to examine the validity of Eq. (1) and that indeed the modified state is a coherent superposition and not mixed. To this end, we perform another experiment, shown in Fig. \ref{second}. We construct a simple Mach-Zehnder interferometer and place two EOMs, with the same modulation frequency, one on each arm. A frequency filter allows only photons with modified frequency to reach the detector. The observed visibility clearly shows that the EOM does not damage the coherence of the photon. In case any information regarding the photon was recorded within the EOM, the visibility of the interference would deteriorate significantly. This additional experiment clearly shows that part amplitude of a single photon is frequency-shifted by the EOM. The photon passing through the EOM is in the superposition state of various frequencies. The classical view that some of the photons are modulated and others are not, is rejected by the observed interference between frequency-shifted components of single photons.

Let us comment briefly on the implication of this result regarding the prospect of counter factual computation (CFC) \cite{cfcn}. CFC is accomplished by putting the computer in one arm of a interferometer so that to prepare it in a superposition of `running' and `not running' states, and then interfering the two histories to obtain the results. The ideas at the basis of CFC have a fundamental disagreement with the TSVF regarding the question of whether the photon passed in a specific path \cite{CFC}. The operational approach we used above, which is supported by the experimental result, adds a practical perspective that can clarify the somewhat philosophical discussion. Let us consider that one could devise a protocol of CFC in which the question of whether or not the computer had run has some operational meaning (we are not aware of such protocol). Using our method it would be possible to analyze the validity of the protocol, even without implementing it in full, a task that might require building a quantum computer.

All the results we present here can be derived using the standard formalism of quantum mechanics. In this derivation, one would have to take into account many details, such as the operation of the EOM, the specific form of the initial spectrum, the technique for measuring the frequency and many more. The cumbersomeness of the calculation would make it extremely difficult to generalize it so it could be applied in a variety of scenarios. On the other hand the TSVF offer a much simpler method. Once the operation of the measurement device is determined, as in Eq. (\ref{meter}), the details of its operation are insignificant. One can simply trace the forward and backward evolving wavefunction and obtain the results directly for any scenario.

\section{V. CONCLUSIONS}
In conclusion, we have performed an experiment to study the past of a particle in a nested Mach-Zehnder interferometer. Using single photons, we recorded information, regarding their path, in the spectrum. We used this information to reconstruct the trajectories of the photons using an operational definition. The results are anomalous trajectories that could be surprising if one attempt to apply a common sense approach to the scenario, but have natural explanation in terms of the TSVF. To achieve this, we developed a new technique for performing weak measurement using the spectrum of a photon.

It would be interesting to see how this scenario can be analyzed within different frameworks and whether other notions related to quantum mechanics can be applied here. The method for weak measurements can be used to devise new experiments or new quantum protocols. Our results demonstrate that single photons passing through EOM are in the superposition states of different frequencies, which may create frequency-encoded photonic qubits for applications in quantum information science \cite{frequency}.\\

{\bf  Acknowledgments}
This work was supported by National Natural Science Foundation of China (Grant Nos. 61327901, 11504362, 11325419, 11654002), the Strategic Priority Research Program (B) of the Chinese Academy of Sciences (Grant No. XDB01030300), the Fundamental Research Funds for the Central Universities (Grant Nos. WK2470000023 and WK2470000024), Key Research Program of Frontier Sciences, The Chinese Academy of Sciences (QYZDY-SSW-SLH003). The work of YK was supported by ERC DM 321031 and VR. YK is grateful to C. Triola for helpful discussions.
Z.-Q.Zhou and X. Liu contributed equally to this work.

\end{document}